\documentclass[onecolumn,showpacs,preprintnumbers,amsmath,amssymb]{revtex4}
\usepackage{bm,amsmath,amssymb,latexsym,mathrsfs,graphicx,enumerate}
\usepackage[mathcal]{euscript}
\usepackage{hyperref}
\usepackage{epsfig}

\usepackage{amsmath}
\usepackage{graphics}
\usepackage{caption}
\DeclareGraphicsExtensions{.bmp,.pdf,.png,.jpg}

\begin{document}

\title{\textbf{Linear guided waves in hyperbolic slab waveguide. Dispersion relations}}
\author{Ekaterina I. Lyashko$^{1}$,  Andrey I. Maimistov$^{1,2}$}
\affiliation{\normalsize \noindent
$^1$: Department of General
Physics, Moscow Institute for Physics
and Technology, Dolgoprudny, Moscow region, 141700 Russia \\
$^2$: Department of Solid State Physics and Nanostructures, National
Nuclear Research University
Moscow Engineering Physics Institute, Moscow, 115409  \\
E-mails: Katerina20490@yandex.ru, ~~aimaimistov@gmail.com \\
}
\date{\today}

\begin{abstract}
\noindent Guided waves modes in a slab waveguide formed from the
isotropic dielectric layer embedded by hyperbolic materials are
investigated. Optical axis is normal to the slab plane. The
dispersion relations for TE and TM waves are found. The differences
between hyperbolic waveguide and conventional one are demonstrated.
In particular, for each TM mode of hyperbolic waveguide there are
two cut-off frequencies and the number of modes is limited. For the
TE and TM modes Poynting vector component along the wave's
propagation axis could be equal to zero.

\end{abstract}

\pacs{42.82.-m, 42.79.Gn, 78.67.Pt}


\maketitle

\date{\today}

\section{Introduction}

\noindent Metamaterials and their optical properties attract grate
attention over the past decade. Metamaterials are artificial
materials which are composed of unit cells far below the size of the
wavelength. These materials can exhibit exotic
electromagnetic properties. The negative refraction is the famous
property of these media
\cite{ShSSch:01,Ch:Wu:Kong:06,Boltasseva:08,Agran:Gart:06,Rautiyan:08,Dolling:07}.
As usually such materials are materials with simultaneously negative
real parts of the dielectric permittivity  and the magnetic
permeability in some frequency region \cite{Elefth:05,Noginov:12}.

It has been known, that the negative refraction can take place in the
anisotropic media
\cite{Podolskiy:05,Hu:07,Makarov:09,Kriegler:10,Smolyani:09}. It
should be pointed, that uniaxial anisotropy is typical property of
the metamaterials. It is important to remark, that an anisotropic
metamaterials can demonstrate negative refraction in one direction
and positive refractions in the orthogonal directions.

Let us assume, that in the uniaxial anisotropic medium coordinate
axes $OX$, $OY$ and $OZ$ are chosen to be equal to the principal
axes of the dielectric permittivity tensor with the principal
values of dielectric permittivity, which satisfy the conditions
$\varepsilon_{xx}=\varepsilon_e$ and
$\varepsilon_{yy}=\varepsilon_{zz}=\varepsilon_o$. The dispersion
relation for extraordinary wave connecting frequency $\omega$ with
Cartesian components of wave vector $\mathbf{k}$, takes the
following form:
\begin{equation}\label{lm:HM:1}
    \frac{k_z^2+k_y^2}{\varepsilon_e(\omega)}+\frac{k_x^2}{\varepsilon_o(\omega)} =
\frac{\omega^2}{c^2}.
\end{equation}
This relation shows, that in the case of either $\varepsilon_e$ or
$\varepsilon_o$ is negative, iso-frequency dispersion surface
(\ref{lm:HM:1}) represents the hyperboloid. The hyperboloid of one
sheet is realized if $\varepsilon_e>0$, $\varepsilon_o<0$ and
hyperboloid of two sheets is obtained if $\varepsilon_e<0$,
$\varepsilon_o>0$
\cite{Wood:06,Noginov:09,XNi:11,Drachev:13,Prashant:14}.
\begin{equation}\label{lm:HM:2}
    \frac{k_z^2+k_y^2}{\varepsilon_e(\omega)}-\frac{k_x^2}{|\varepsilon_o(\omega)|} =
\frac{\omega^2}{c^2}, \qquad \frac{k_x^2}{\varepsilon_o(\omega)}
-\frac{k_z^2+k_y^2}{|\varepsilon_e(\omega)|} = \frac{\omega^2}{c^2},
\end{equation}
These anisotropic materials are referred to as hyperbolic materials.

The hyperbolic materials can be fabricated as a multilayer structure
consisting of alternating metallic and dielectric layers
\cite{Wood:06,Kivshar:13,Othman:13}, or as a nanowire structure
consisting of metallic nanorods embedded in a dielectric host
\cite{Lu:Sridhar:08,Silveirinha:09,Noginov:09}.

Iso-frequency dispersion surfaces (\ref{lm:HM:2}) allow the
infinitely large wave vectors. It results in the different effects,
among which are the Purcell enhancement of the spontaneous emission
rate in hyperbolic metamaterials
\cite{Poddubny:12,Poddubny:13,Ward:13,Ferrari:14} and the
subwavelength resolution effect \cite{Wood:06,Benedict:13}.

The optical phenomena on interface between conventional dielectric
and hyperbolic metamaterial has attracted attention. The
surface waves were studied in \cite{Zapata:13}.  The extremely large
Goos–-H\"{a}nchen shift has been studied in some details in
\cite{JingZhao:13}. Plasmonic planar waveguide cladded by hyperbolic
metamaterials (Fig.1) was proposed and investigated in
\cite{Kildishev:14,Kildishev:15}.

The purpose of this paper is to investigate the dispersion
properties of the linear guided waves in a planar waveguide
previously studied in \cite{Kildishev:14}. Unlike
\cite{Kildishev:14} the guided waves localized in the dielectric
core will be considered here instead of surface waves. The
anisotropy axes of substrate and cladding layers are directed along
$OX$ axis that is normal to the interface (Fig.1). In the planar
geometry, as is known \cite{Tamir:78}, the Maxwell equations can be
separated into two uncoupled systems of equations, which describe
the propagation of the waves having a different polarization. These
waves are refereed to as TE and TM waves. Analysis of the guided TE
and TM waves will be done independently. In both cases the
expression for dispersion relation connecting effective waveguide
index with frequency will be obtained. The symmetrical waveguide
will considered in details.

\section{Electric and magnetic field distributions for guided waves}

\noindent Let us consider a slab waveguide. We assume, that the
material of the waveguide core is nonmagnetic $\mu_i =1$ and has an
isotropic permittivity $\varepsilon_i$. The core thickness is $h$.
The dielectric core is cladded by the uniaxial hyperbolic
metamaterials which are characterized by symmetric dielectric tensor
with the principal dielectric constants $\varepsilon_o^{(1)}$,
$\varepsilon_e^{(1)}$, $\varepsilon_o^{(3)}$, $\varepsilon_e^{(3)}$
and the magnetic permeabilities $\mu_1$ and $\mu_3$. All permeabilities
are assuming to be positive. The anisotropy axes are aligned with a
unit normal vector to the interface, i.e., along the $OX$ direction.
(Fig.1). Axes $OY$ and $OZ$ are parallel to interface. Axis $OZ$ is directed along the wave propagation. In this case the Maxwell
equations are invariant under the shifting along $OY$ axis. Thus, the
strengths of the electric and magnetic fields of the guided wave are
independent of variable $y$. From it follows, that the Maxwell equations
are splitting into two independent systems of equations describing
the TE and TM waves \cite{Tamir:78}.

\begin{figure}[h]
    \center{\includegraphics[scale=0.45]{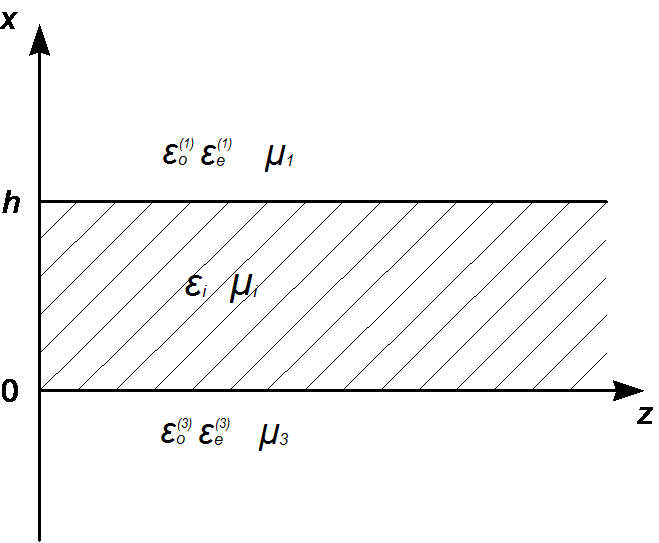}}
    \caption{ A schematic illustration of the
hyperbolic waveguide}
\end{figure}

The TE wave is defined by the tangent component of the electric field
vector $E_y$ and by two components of magnetic field: $H_x$ and
$H_z$. These values are assumed to be harmonic functions of the time:
$\exp(i\omega t)$. The wave equation for complex amplitude
 $E=E_y(x,z,\omega)$ looks like:
$$
\frac{\partial^2 E }{\partial x^2}+\frac{\partial^2 E }{\partial
z^2} +k_0^2\varepsilon_o(x)\mu(x) E =0,
$$
where $k_0=\omega/c$, $\omega$ is frequency of radiation. The
principal dielectric constants and magnetic permeabilities are
piecewise functions (Fig.1):
$$
\varepsilon_o(x)=\left\{
\begin{array}{ccc}
  \varepsilon_o^{(1)} & x<0, \\
  \varepsilon_i & 0 \leq x \leq h, \\
  \varepsilon_o^{(3)} & x>h, \\
\end{array}\right.\quad \varepsilon_e(x)=\left\{
\begin{array}{ccc}
  \varepsilon_e^{(1)} & x<0, \\
  \varepsilon_i & 0 \leq x \leq h, \\
  \varepsilon_e^{(3)} & x>h, \\
\end{array}\right. \quad \mu(x)=\left\{
\begin{array}{ccc}
  \mu_1 & x<0, \\
  \mu_i & 0 \leq x \leq h, \\
  \mu_3 & x>h, \\
\end{array}\right.
$$
The components of magnetic field can be found from the following relations:
\begin{equation}\label{eq:LM:Hx:Hz:1}
 H_x=\frac{i}{k_0\mu(x)} \frac{\partial E }{\partial z}, \quad
 H_z=-\frac{i}{k_0\mu(x)} \frac{\partial E }{\partial x}.
\end{equation}

The TE wave is ordinary one according to choosing of the direction
of anisotropy axis. In the case of $\varepsilon_o>0$ the problem is
reduced to a familiar case. However, one can expect the interesting
result for hyperbolic material with $\varepsilon_o<0$.

The electric field can be represented in the form $E(x,z) =
\tilde{E}(x)\exp(i \beta z)$, because the waveguide is homogeneous
along $OZ$, where parameter $\beta$ is the propagation constant.
This parameter is similar to the wave number in the case of
homogeneous medium. Solutions of the wave equation should be found
with taking into account the boundary conditions $\mathbf{E} \to 0$,
$\mathbf{H} \to 0$ at $|x|\to\infty$. The solution can be obtained
by standard procedure \cite{Tamir:78}. Distribution of the electric
field is given by the following equations
\begin{eqnarray}
  &x<0 :& E^{(1)}= Ae^{px+i\beta z}+c.c.,         \nonumber \\
  &0\leq x \leq h :& E^{(2)}= A\left[\cos(\kappa x)+\xi_p\sin(\kappa x)\right]e^{i\beta z} +c.c., \label{eq:LM:o:TE:1} \\
  &x>h :& E^{(3)} = A\left[\cos(\kappa h)+\xi_p\sin(\kappa h)\right]e^{-q(x-h)}e^{i\beta z}+c.c.\nonumber
\end{eqnarray}
where following parameters are used
$$p^2 = \beta^2 +
k_0^2\mu_1|\varepsilon_o^{(1)}|,\quad q^2 = \beta^2 +
k_0^2\mu_3|\varepsilon_o^{(3)}|,\quad \kappa^2
=k_0^2\mu_i\varepsilon_i-\beta^2.
$$
These parameters can be used to define the Goos–-H\"{a}nchen phase
shifts $\phi_{q}$ and $\phi_{p}$:
$$\xi_q = -\tan(\phi_{q}/2)=
\frac{q\mu_i} {\kappa\mu_3}, \quad \xi_p = -\tan(\phi_{p}/2) =
\frac{p\mu_i} {\kappa\mu_1}.$$ Normalized amplitude $A$ of the
electrical field at $x=0$ is arbitrary.

TM wave is determined by the component of a magnetic field $H_y$ and
an electric field components $E_x$, $E_z$. The wave equation for magnetic field
$H=H_y(x,z,\omega)$ is
\begin{equation}\label{eq:hip:lin:hyp:H-wave}
    \frac{1}{\varepsilon_e(x)}\frac{\partial^2 H}{\partial z^2} +
    \frac{\partial}{\partial x}\left(\frac{1}{\varepsilon_o(x)}\frac{\partial H}{\partial x}\right) +k_0^2\mu(x) H
    = 0.
\end{equation}
The electric field components can be found from the following
expressions:
\begin{equation}\label{eq:LM:Ex:Ez:1}
 E_x=-\frac{i}{k_0\varepsilon_e(x)} \frac{\partial H }{\partial z}, \quad
 E_z=\frac{i}{k_0\varepsilon_o(x)} \frac{\partial H }{\partial x}.
\end{equation}
The principal dielectric constants and permeability are piecewise
functions previously defined in the case of TE waves.

A consideration of the equation (\ref{eq:hip:lin:hyp:H-wave}) shows,
that if $\varepsilon_e^{(a)} <0$ and $\varepsilon_o^{(a)}>0$
($a=1,3$), then there is no solution of this equation decreasing at
$|x|$ tends to infinity. Hence, there are no wave localized in
waveguide. Alternatively, if $\varepsilon_e^{(a)}
>0$ and $\varepsilon_o^{(a)}<0$ then under conditions:
\begin{equation}\label{eq:LM:TM:7}
 k_0^2\mu_1\varepsilon_e^{(1)}>\beta^2,\quad
 k_0^2\mu_3\varepsilon_e^{(3)}>\beta^2
\end{equation}
equation (\ref{eq:hip:lin:hyp:H-wave}) admits a solution describing
wave confinement in this waveguide. The magnetic field distribution
is written as
\begin{eqnarray}
  &x<0 :& H^{(1)}= Ae^{px+i\beta z}+c.c.,         \nonumber \\
  &0\leq x\leq h :& H^{(2)}= A\left[\cos(\kappa x)-\xi_p\sin(\kappa x)\right]e^{i\beta z} +c.c., \label{eq:LM:distr:H:TM:1} \\
  &x>h :& H^{(3)} = A\left[\cos(\kappa h)-\xi_p\sin(\kappa h)\right]e^{-q(x-h)}e^{i\beta z}+c.c.\nonumber
\end{eqnarray}
In these expressions following parameters
$$p^2= k_0^2\mu_1|\varepsilon_o^{(1)}| -
  \frac{|\varepsilon_o^{(1)}|}{\varepsilon_e^{(1)}}\beta^2,\quad q^2= k_0^2\mu_3|\varepsilon_0^{(3)}| -
  \frac{|\varepsilon_o^{(3)}|}{\varepsilon_e^{(3)}}\beta^2,\quad \kappa^2
=k_0^2\mu_i\varepsilon_i-\beta^2,
$$
are used. The Goos–H\"{a}nchen phase shifts $\phi_{q}$ and
$\phi_{p}$ are defined by the following expressions:
$$\xi_q = \tan(\phi_{q}/2)=
\frac{q\varepsilon_i} {\kappa|\varepsilon_o^{(3)}|}, \quad \xi_p =
\tan(\phi_{p}/2) = \frac{p\varepsilon_i}
{\kappa|\varepsilon_o^{(1)}|}.$$

\section{Dispersion relations}

\noindent Taking into account decreasing of the magnetic and
electric fields in the limit $|x|\to \infty$, the Maxwell equations
solutions describe the waves which are confined by the waveguide. It
is necessary to distinguish the coupled surface waves and the
waveguide modes. In a linear waveguide the amplitude of surface wave
takes the maximum at interface. In the case of the dielectric
waveguide cladded by metal this coupled surface wave is said to be
the plasmon-polariton wave. Guided plasmon-polariton wave in the
dielectric waveguide arranged from a hyperbolic metamaterial has
been proposed and studied in \cite{Kildishev:14,Kildishev:15}. The
dispersion relation of this plasmon-polariton wave can be obtained
by changing parameter $\kappa^2$ to $\kappa^2
=\beta^2-k_0^2\mu_i\varepsilon_i$. However, other than these, there
are a number of waves localized in the dielectric core that are
designated as guided modes or waveguide modes \cite{Tamir:78}.

For all guided waves the dispersion relation exists as
expression connecting the propagation constant $\beta$ and frequency
$\omega$. The dispersion relations of the guided waves under
consideration can be determined by using the continuity conditions of
an electric and magnetic field on the interferences. It is suitable to
get the dispersion relations for TE and TM wave separately.

\subsection{Case of TE wave}

\noindent Distribution of the magnetic field in waveguide is
derivable from (\ref{eq:LM:o:TE:1}) with taking into account the
expressions (\ref{eq:LM:Hx:Hz:1}). The continuity conditions for
tangent components of both electric and magnetic field vectors
result in following relation:
$$
e^{2i\kappa h}\left(\frac{1-i\xi_q}{1 +i\xi_q}\right)
\left(\frac{1-i\xi_p}{1+i\xi_p}\right) =1.
$$
Using the expression for the Goos–-H\"{a}nchen phase shift one can
write the dispersion relation in form:
\begin{equation}\label{eq:hip:lin:hyp:TE:disp:rel}
    2\kappa h + \phi_p+\phi_q = 2\pi m, \quad m = 0,~1,~2,\ldots
\end{equation}
If the effective index of refraction $n_{ef}$ is defined according
to formula $\beta = k_0n_{ef}$, than equation
(\ref{eq:hip:lin:hyp:TE:disp:rel}) can be written as:
$$
hk_0\sqrt{n^2_i-n_{ef}^2}=
\arctan\left(\frac{\mu_i}{\mu_1}\sqrt{\frac{n_1^2+n_{ef}^2}{n_i^2-n_{ef}^2}}\right)
+
\arctan\left(\frac{\mu_i}{\mu_3}\sqrt{\frac{n_3^2+n_{ef}^2}{n_i^2-n_{ef}^2}}\right)
+\pi m.
$$
Here the indexes of refraction $ n_1^2 = \mu_1
|\varepsilon_o^{(1)}|, ~~n_3^2= \mu_3 |\varepsilon_o^{(3)}|,
~~n_i^2=\mu_i\varepsilon_i $ are introduced.

The dispersion relation shows that the effective index of
refraction is limited by the condition $0 \leq n_{ef}^2 < n_i^2$. In
the case of all dielectric waveguide the similar limitation appears
as $\max (n_1^2, n_3^2) \leq n_{ef}^2 < n_i^2$. Difference between
these inequalities is due to the fact that 
in dielectric waveguide embedded in a hyperbolic media 
the total internal reflection takes place at any incident angle. As
in the case of dielectric waveguide cladded by metal.

Analysis of the dispersion relations will be performed for the case
of symmetric waveguide, where $n_1^2 =n_3^2 $. Then the dispersion
relation takes the form:
\begin{equation}\label{eq:LM:TE:5}
    k_0h\sqrt{n_i^2-n_{ef}^2} =2\arctan
    \left(\frac{\mu_i}{\mu_1}\sqrt{\frac{n_1^2+n_{ef}^2}{n_i^2-n_{ef}^2}}\right) + \pi m.
\end{equation}
Equation (\ref{eq:LM:TE:5}) can be rewritten in the normalized form. The
parameter $b$ is introduced by the following equation
$n_1^2+n_{ef}^2=b\Delta$, where $\Delta=n_1^2+n_{i}^2$. The
normalized waveguide thickness $V$ is introduced by formula  $V=
k_0h\sqrt{n_i^2+n_{1}^2}$. Then the relation (\ref{eq:LM:TE:5})
takes the form:
\begin{equation}\label{eq:LM:TE:bvsV}
   V\sqrt{1-b} =2\arctan
    \left(\frac{\mu_i}{\mu_1}\sqrt{\frac{b}{1-b}}\right) + \pi m.
\end{equation}

\begin{figure}[h!]
    \center{\includegraphics[scale=0.60]{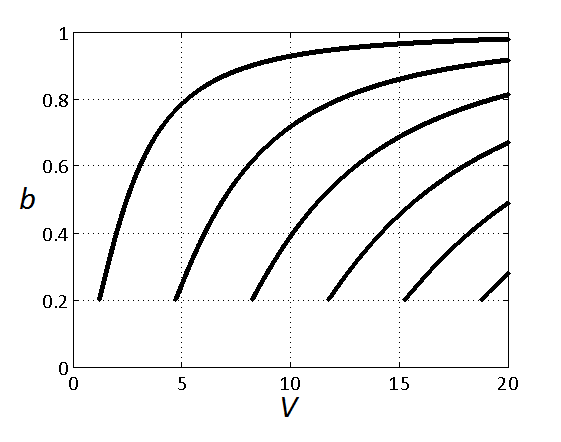}}
    \caption{Dispersion curves for TE modes of the hyperbolic waveguide.}
\end{figure}

This equation defines the function $b(V, m)$ that is implicit
dependence of the normalized effective index of refraction $b$ on
normalized waveguide thickness $V$. In the case under consideration
$b$ is in the interval $[b_0, 1)$, where $b_0=n_1^2/(n_1^2+n_i^2)$. The
plots of $b(V, m)$ vs $V$ are presented on Fig.2 at
$\mu_i/\mu_2=1.2$ and $b_0=0.2$. The fact that $b_0> 0$ means that
zero-mode TE$_0$ has non zero cut-off frequency $V_{c0}$. Substitution
$b=b_0$ into equation (\ref{eq:LM:TE:bvsV}) results in
\begin{equation}\label{eq:LM:TE:V:catof}
   V_{c0} =2\sqrt{1+\frac{n_1^2}{n_i^2}}\arctan
    \left(\frac{\mu_i n_1}{\mu_1 n_i}\right).
\end{equation}
For conventional dielectric waveguide cut-off frequency $V_{c0}$ is
zero.

\subsection{Case of TM wave}

\noindent By the use of (\ref{eq:LM:distr:H:TM:1}) and
(\ref{eq:LM:Ex:Ez:1}) the electric field strength cab be found. The
continuity conditions for tangent components of both electric and
magnetic field vectors lead to the following dispersion relation for TM
waves
$$
e^{2i\kappa h}\left(\frac{1+i\xi_q}{1-i\xi_q}\right)
\left(\frac{1+i\xi_p}{1-i\xi_p}\right) =1.
$$
Using the Goos--H\"{a}nchen phase shifts the dispersion relation can
be written as :
\begin{equation}\label{eq:hip:lin:hyp:TM:disp:rel}
    2\kappa h + \phi_p+\phi_q = 2\pi m, \quad m = 0,~1,~2,\ldots
\end{equation}
In terms of initial variables the relation
(\ref{eq:hip:lin:hyp:TM:disp:rel}) takes the form:
$$ h\sqrt{k_0^2 (\mu_i\varepsilon_i - n^2_{ef})}=-
\arctan \sqrt{
\frac{\varepsilon_i^2}{|\varepsilon_o^{(3)}|\varepsilon_e^{(3)}}
\left(\frac{\varepsilon_e^{(3)}-n^2_{ef}}
{\varepsilon_i-n^2_{ef}}\right) } - \arctan\sqrt{
\frac{\varepsilon_i^2}{|\varepsilon_o^{(1)}|\varepsilon_e^{(1)}}
\left(\frac{\varepsilon_e^{(1)}-n^2_{ef}}
{\varepsilon_i-n^2_{ef}}\right) } +\pi m.
$$

Let us consider only symmetrical waveguide, where $n_1^2 =n_3^2 $.
The dispersion relation can be rewritten as
\begin{equation}\label{eq:LM:TM:9}
    hk_0\sqrt{n^2_i -n_{ef}^2} =-2\arctan
    \left[
    \sqrt{\left(\frac{\varepsilon_i^2}{|\varepsilon_o^{(1)}|\varepsilon_e^{(1)}}\right)
    \frac{n_e^2-n_{ef}^2}
    {n_i^2-n_{ef}^2}}\right] + \pi m,
\end{equation}
where the effective indexes of refraction $n_i$ and $n_e$ are used.
These parameters are defined by correlation
$n_i^2=\mu_i\varepsilon_i$ for isotropic dielectric and
$n_e^2=\mu_1\varepsilon_e$ for extraordinary wave in the hyperbolic
media.

The condition for effective index
$$0\leq n_{ef}^2<n_i^2, \quad 0\leq n_{ef}^2 \leq n_e^2 $$ follows from the equation (\ref{eq:LM:TM:9}).
In the case of a convenient dielectric medium this condition appears
as $n_{c} \leq n_{ef}<n_i$, where $n_{c}$ is the refraction index of
the substrate or cladding layer.

Equation (\ref{eq:LM:TM:9}) can be rewritten in terms of the uniform
variables ($b$, $V$) by using following correlations:
$n_e^2-n_{ef}^2= b\Delta
>0$, where $\Delta=n_i^2-n_{e}^2$, and normalized waveguide thickness $V$
is introduced as $V= k_0h\sqrt{n_i^2-n_{e}^2}$. $b$ is normalized
effective index of refraction of waveguide. It results in the
dispersion relation in following form:
\begin{equation}\label{eq:LM:TM:10}
   V\sqrt{1+b} =-2\arctan
    \left[\sqrt{\frac{\varepsilon_i^2}{|\varepsilon_o^{(1)}|\varepsilon_e^{(1)}}
    \left(\frac{b}{1+b}\right)}\right] + \pi m.
\end{equation}

In the case under consideration the parameter $b$ is in interval
 $[0, b_0]$, where $b_0=n_e^2/(n_i^2-n_e^2)$.

The dispersion curves corresponding to equation (\ref{eq:LM:TM:10})
are represented in Fig.3. The assumption
$\varepsilon_i^2/(|\varepsilon_o^{(1)}|\varepsilon_e^{(1)})=1.2$,
$b_0=2$ is hold. For comparison the dispersion curves corresponding
all dielectric waveguide are shown in Fig.4.
$$
V\sqrt{1-b} =2\arctan\left(\sqrt{u\frac{b}{1-b}}\right) +\pi m,
\quad m=0,1,2,\dots $$ where $u=
\varepsilon_i^2/(\varepsilon_o\varepsilon_e)$. The curves in Fig. 4
were obtained at $u=1.2$.

\begin{figure}[h!]
    \center{\includegraphics[scale=0.75]{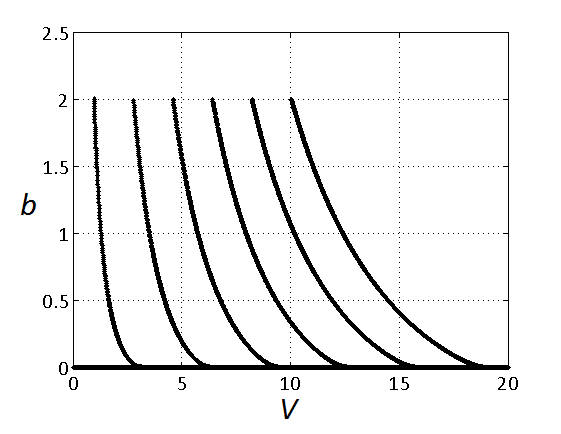}}
    \caption{Dispersion curves for TM modes of the hyperbolic waveguide.}
\end{figure}

\begin{figure}[h!]
    \center{\includegraphics[scale=0.60]{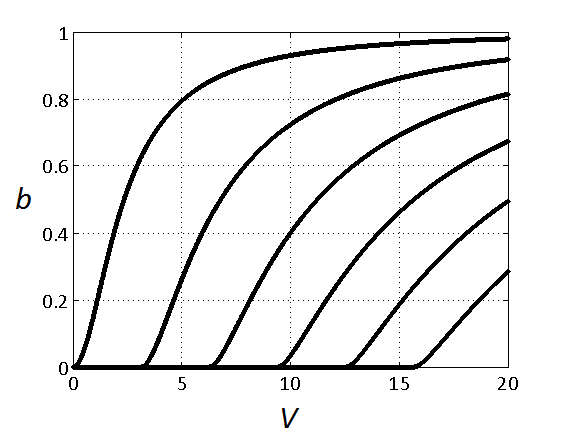}}
    \caption{Dispersion curves for TM modes of the conventional dielectric waveguide.}
\end{figure}

Figures show, that in the case of TM wave the number of guided modes
of the hyperbolic waveguide is limited. As the dielectric core
thickness $h$ increases, one mode disappears but other mode appears.
Furthermore, the zero-mode ($m=0$) in this waveguide is absent. For
an all dielectric waveguide (Fig.4) the number of guided modes
increases with core thickness $h$.

Thus, for the each TM mode of the hyperbolic waveguide two cut-off
frequencies exist: $b(V_{c~m}^{(2)})=b_0$ and $b(V_{c~m}^{(1)})=0$.
For each TM mode of conventional dielectric waveguide only
single cut-off frequency exists $V_{c~m}^{(1)}$.

\subsection{Poynting vector of the guided waves in the hyperbolic waveguide}

\noindent The Poynting vector defines density of the radiation
energy flux and direction of wave's energy propagation. It is
instructive to consider an averaged projection of the Poynting
vector along the $OZ$ axis. For TE wave it can be found from the
equation
\begin{equation}\label{eq:Point:TE}
    \langle S_z \rangle =-\frac{c}{16\pi}(E_y^*H_x+E_yH_x^*),
\end{equation}
and for TM wave it follows from the relation
\begin{equation}\label{eq:Point:TM}
\langle S_z \rangle =\frac{c}{16\pi}(E_x^*H_y+E_xH_y^*).
\end{equation}
The relations (\ref{eq:Point:TE}) and (\ref{eq:Point:TM}) can be
rewritten with taking into account equation (\ref{eq:LM:Hx:Hz:1})
for the TE wave and equation (\ref{eq:LM:Ex:Ez:1}) for the TM wave
as
\begin{equation}\label{eq:Point}
\langle S_z \rangle_{TE}=\frac{c}{8\pi}\frac{n_{ef}}{\mu_j}|E_y|^2,
\qquad \langle S_z \rangle_{TM}=
\frac{c}{8\pi}\frac{n_{ef}}{\varepsilon_j}|H_y|^2,
\end{equation}
where $j=1, i, 3$ for $\mu_j$ and $j=e^{(1)}, i, e^{(3)}$ for
$\varepsilon_j$ depending on the layer under consideration.

As was obtained in the previous two subsections, the effective index
of refraction $n_{ef}$ for TE and TM guided modes in the hyperbolic
waveguide can achieve null value. In these cases the averaged energy
flux along the guided wave propagation axis $OZ$ will be zero. Thus,
the effect of slowing light in the waveguide takes place in these
cases.

\section{Conclusion}

\noindent The special kind of the hyperbolic slab waveguide is
considered here. In the case of $\varepsilon_o<0$ and
$\varepsilon_e>0$ the modes of directed waves were found. The TE
wave is ordinary wave, but TM wave is extraordinary in the waveguide
under consideration. If $\varepsilon_o>0$ and $\varepsilon_e<0$  TM
wave is not confined, and TE guided waves are identical with waves
in a conventional waveguide. It is the reason to study the case of
hyperbolic waveguide with $\varepsilon_o<0$ and $\varepsilon_e>0$.

The effective index of refraction for TE wave obeys following
inequality $0 \leq n_{ef}<n_i$, where $n_i$ is the index of
refraction of the waveguide's core. In the case of TM wave the
effective index of refraction varies within the limits $0 \leq
n_{ef} \leq n_e$, where $n_e$ is the index of refraction of the
extraordinary wave in hyperbolic medium. (It is assumed that $n_e<
n_i$.) Thus, in both cases the effective index of refraction can be
equal to zero, that leads to great phase velocity of the wave and
zero Poynting vector component along the wave's propagation axis in
the hyperbolic waveguide discussed above. Hence the light wave can
be slowed down in this waveguide. In the case of conventional
(elliptic) anisotropic dielectric waveguide the effective index of
refraction lies in the range $n_e \leq n_{ef}<n_i$ for TM wave, and
in the range $n_o \leq n_{ef}<n_i$ for TE wave.

The dispersion relations for the case of TE and TM waves are
derived. It was shown that for the TM wave the number of guided
modes is limited. Each of these modes have two cut-off frequencies.
One of them corresponds to mode appearance, another corresponds to
mode disappearance. There is region of parameters in which the only
single mode exists in this waveguide. It is worth noting that this
phenomenon is unavailable in the case of a conventional waveguide.
Usually the number of modes increases with core thickness, and only
single cut-off frequency exists.

\section*{ Acknowledgement}

We are grateful to Prof. I. Gabitov and Dr. C. Bayun for
enlightening discussions. This investigation is funded by Russian
Science Foundation (project 14-22-00098).

\end{document}